\documentclass[showpacs,10pt,twocolumn,prl]{revtex4-1}

\usepackage{amsmath}
\usepackage{amssymb}
\usepackage{graphicx}
\usepackage{amssymb}
\usepackage{graphics}
\usepackage{epsfig}
\usepackage{CJK}
\usepackage{color}
\usepackage{soul}

\begin{document}

\begin{CJK*}{GBK}{Song}
\title{Three-dimensional ferromagnetism and magnetotransport in van der Waals Mn-intercalated tantalum disufide}
\author{Yu Liu,$^{1,*}$ Zhixiang Hu,$^{1,2}$ Eli Stavitski,$^{3}$, Klaus Attenkofer,$^{3,\dag}$ and C. Petrovic$^{1,2}$}
\affiliation{$^{1}$Condensed Matter Physics and Materials Science Department, Brookhaven National Laboratory, Upton, New York 11973, USA\\
$^{2}$Materials Science and Chemical Engineering Department, Stony Brook University, Stony Brook, New York 11790, USA\\
$^{3}$National Synchrotron Light Source II, Brookhaven National Laboratory, Upton, New York 11973, USA}
\date{\today}

\begin{abstract}
Van der Waals (vdW) ferromagnets are an important class of materials for spintronics applications. The recent discovery of atomically vdW magnets CrI$_3$ and Cr$_2$Ge$_2$Te$_6$ has triggered a renaissance in the area of two-dimensional (2D) magnetism. Herein we systematically studied 2H-Mn$_{0.28}$TaS$_2$ single crystal, a 2D vdW ferromagnet with $T_c$ $\sim$ 82.3 K and a large in-plane magnetic anisotropy. Mn $K$-edge x-ray absorption spectroscopy was measured to provide information on its electronic state and local atomic environment. The detailed magnetic isotherms measured in the vicinity of $T_c$ indicates that the spin coupling inside 2H-Mn$_{0.28}$TaS$_2$ is of a three-dimensional (3D) Heisenberg-type coupled with the attractive long-range interaction between spins that decay as $J(r)\approx r^{-4.85}$. Both resistivity $\rho(T)$ and thermopower $S(T)$ exhibit anomalies near $T_c$, confirming that the hole-type transport carriers strongly interact with local moments. An unusual angle-dependent magnetoresistance is further observed, suggesting a possible field-induced novel magnetic structure.
\end{abstract}
\maketitle
\end{CJK*}

\section{INTRODUCTION}

Layered vdW materials attract great interest due to their remarkable physical properties and applications in electronic/spintronic devices \cite{Geim,Butler,Bhimanapati,Novoselov}. Recently, several notable intrinsic vdW magnets have been observed in mono-/few-layer of FePS$_3$, Cr$_2$Ge$_2$Te$_6$, CrI$_3$, Fe$_3$GeTe$_2$, VSe$_2$, and MnSe$_2$, of high interest in spintronic and in fundamental 2D magnetism alike \cite{Lee,Huang,Gong,Deng,Bonilla,Hara}.

Magnetic critical behavior gives insight into the nature of magnetic interactions, correlation length, spin dimensionality, and the spatial decay of correlation function at criticality \cite{Fisher0,Stanley0}. Three principal spin Hamiltonians are reported in MPS$_3$ (M = transition metal): 2D Heisenberg type in MnPS$_3$, 2D XY type in NiPS$_3$, and 2D Ising type in FePS$_3$ \cite{Joy}. Moreover, 2D Ising-like critical behavior is observed in bulk Cr$_2$Si$_2$Te$_6$ and Cr$_2$Ge$_2$Te$_6$ \cite{BJLiu0,YuLiu0,WLiu0,GTLin0}, however 3D critical exponents are found in bulk Fe$_3$GeTe$_2$, CrI$_3$, and Mn$_3$Si$_2$Te$_6$, as a result of the different strengths of interlayer coupling \cite{YuLiu1,BJLiu1,YuLiu2,GTLin1,YuLiu3}. The magnetic properties can further be tuned by thickness variation, as for example a crossover 3D-2D behavior in thickness-reduced CrI$_3$ and Fe$_3$GeTe$_2$ \cite{YuLiu4,Fei}. In addition, intercalated transition metal dichalcogenides commonly feature 3$d$ atoms in the vdW gap and host diverse magnetic orders \cite{Parkin,Friend0,Guo,Friend,Zhang}. A notable example is 2H-Fe$_{0.25}$TaS$_2$, the only member that exhibit ferromagnetic (FM) order with a strong uniaxial anisotropy, similar to recently investigated 2D vdW magnets.  2H-Fe$_{0.25}$TaS$_2$ features large magnetocrystalline anisotropy, magnetoresistance (MR), and sharp switching in magnetization \cite{Morosan,Hardy,Chen}. 2H-Mn$_x$TaS$_2$ has been found to be FM but with an easy-plane anisotropy \cite{Hinode,Onuki,Cussen,Li,Shand,Zhang1}.

In this work, we systematically studied the magnetic critical behavior and magnetotransport properties of 2H-Mn$_{0.28}$TaS$_2$ single crystal, with relatively high $T_c$ of 82.3 K and large in-plane anisotropy. Critical exponents $\beta = 0.57(1)$, $\gamma = 1.27(3)$, and $\delta = 3.51(1)$ are close to those calculated via renormalization group approach for a 3D Heisenberg model coupled with attractive long-range interaction between spins that decay as $J(r)\approx r^{-4.85}$. 2H-Mn$_{0.28}$TaS$_2$ shows a metallic behavior in resistivity $\rho(T)$ and positive values of thermopower $S(T)$ with dominant hole-type carriers, in which obvious anomalies were observed corresponding to the magnetic transition. The observed unusual angle-dependent MR suggests a change of magnetic state in magnetic field.

\section{EXPERIMENTAL DETAILS}

\begin{figure*}
\centerline{\includegraphics[scale=1]{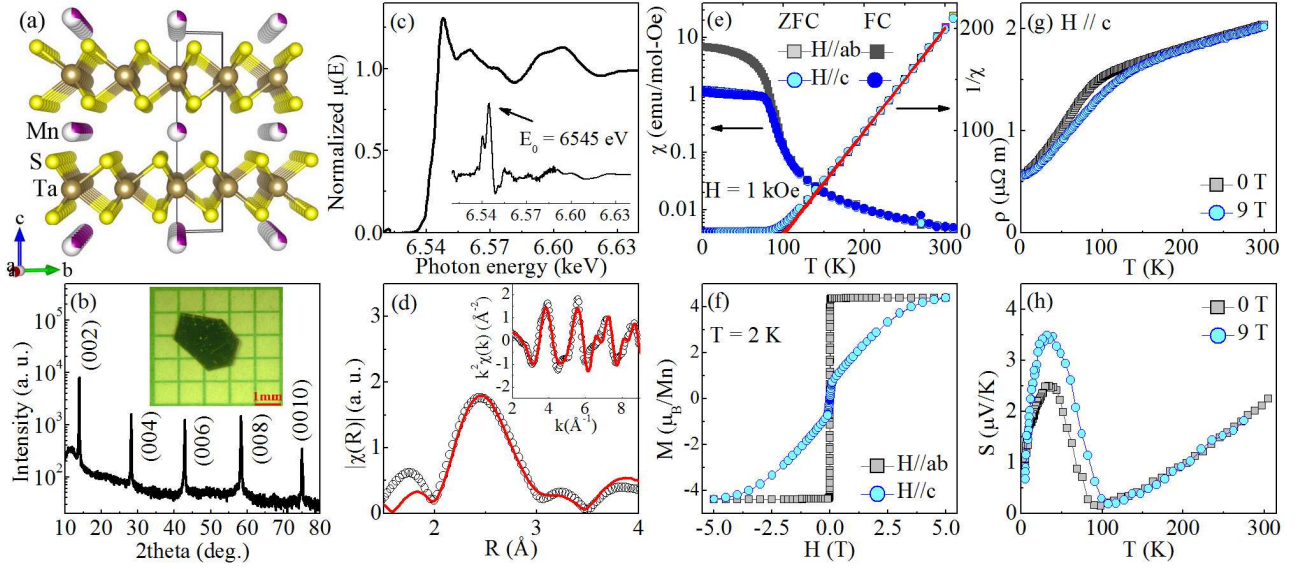}}
\caption{(Color online) (a) Crystal structure and (b) x-ray diffraction (XRD) pattern in logarithmic scale of 2H-Mn$_{0.28}$TaS$_2$. Inset in (b) shows the representative single crystal. (c) Normalized Mn $K$-edge x-ray absorption near edge structure (XANES) spectra. Inset shows the derivative $\mu(E)$ curve. (d) Fourier transform magnitudes of the extended x-ray absorption fine structure (EXAFS) oscillations (symbols) for Mn $K$-edge with the phase shift correction. The model fits are shown as solid lines. Inset shows the corresponding filtered EXAFS (symbols) with $k$-space model fits (solid lines). (e) Temperature-dependent magnetic susceptibility $\chi(T)$ {\color{blue}(left axis)} and inverse susceptibility $1/\chi(T)$ {\color{blue}(right axis)} with zero-field cooling (ZFC) and field cooling (FC) processes for $\mathbf{H} \parallel \mathbf{c}$-axis and $\mathbf{H} \parallel \mathbf{ab}$-plane, respectively. The solid straight lines show the Curie-Weiss law fits from 150 to 300 K {\color{blue}(right axis)}. (f) Field-dependent magnetization at 2 K for $\mathbf{H} \parallel \mathbf{c}$-axis and $\mathbf{H} \parallel \mathbf{ab}$-plane, respectively. Temperature dependence of in-plane (g) resistivity $\rho(T)$ and (h) thermopower $S(T)$ in 0 and 9 T.}
\label{XRD}
\end{figure*}

The single crystals were fabricated by chemical vapor transport method with iodine agent. A mixture of Mn, Ta, and S powder with a nominal mole ratio of 0.5 $:$ 1 $:$ 2 was sealed in an evacuated quartz tube and then heated for two weeks in a two-zone furnace with the source zone temperature of 1000 $^\circ$C and the growth zone temperature of 900 $^\circ$C, respectively. The obtained single crystals are hexagonal shape with typical dimensions as $2\times2\times0.5$ mm$^3$. The average stoichiometry was determined by examination of multiple points on fresh surface and checked by multiple samples from the same batch using energy-dispersive x-ray spectroscopy in a JEOL LSM-6500 scanning electron microscope. The actual element ratio is Mn$_{0.28(1)}$TaS$_{1.88(2)}$ with Ta fixed to 1.0, which is referred to as Mn$_{0.28}$TaS$_2$ throughout this paper due to $\sim 5\%$ sulfur deficiency is normally observed in transition-metal disulfide. X-ray diffraction (XRD) pattern was acquired on a Rigaku Miniflex powder diffractometer with Cu $K_{\alpha}$ ($\lambda=0.15418$ nm). X-ray absorption spectroscopy was measured at 8-ID beamline of the National Synchrotron Light Source II (NSLS II) at Brookhaven National Laboratory (BNL) in the fluorescence mode. The x-ray absorption near edge structure (XANES) and extended X-ray absorption fine structure (EXAFS) spectra were processed using the Athena software package. The EXAFS signal, $\chi(k)$, was weighed by $k^2$ to emphasize the high-energy oscillation and then Fourier-transformed in $k$ range from 2 to 10 {\AA}$^{-1}$ to analyze the data in $R$ space. The electrical, thermal transport, and magnetization were measured in quantum design PPMS-9 and MPMS-5 instruments.

\section{RESULTS AND DISCUSSIONS}

\begin{figure*}
\centerline{\includegraphics[scale=1]{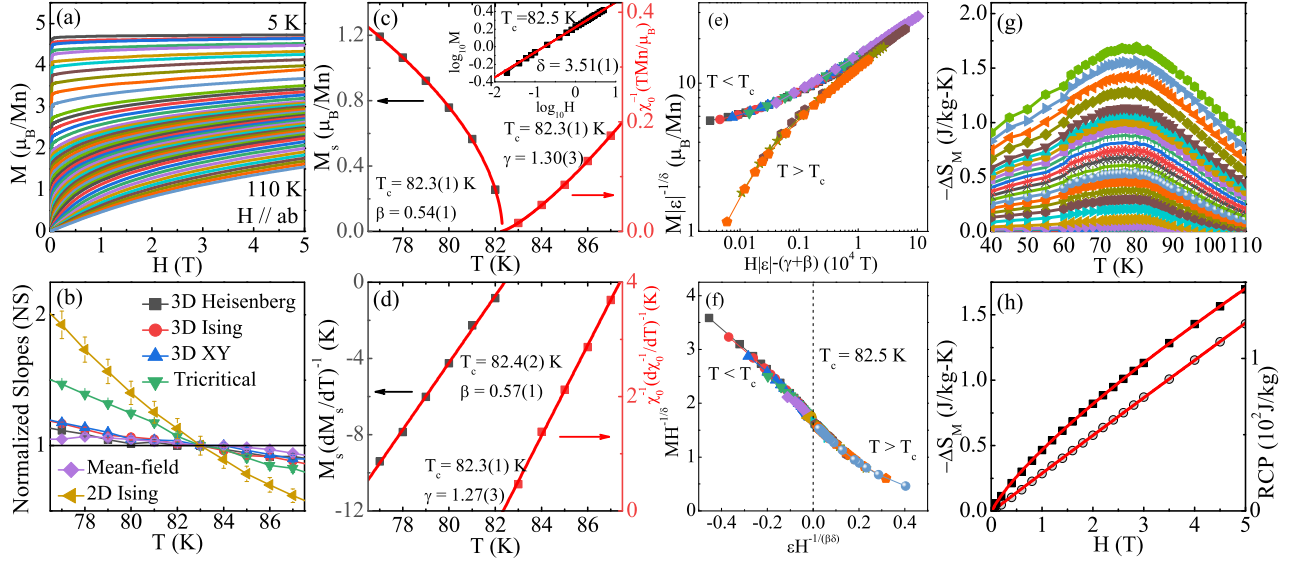}}
\caption{(Color online) (a) Typical initial isothermal magnetization curves measured with $\mathbf{H} \parallel \mathbf{ab}$-plane from 5 to 110 K with temperature steps (5-55 K: 5K; 60-68 K: 2K; 70-89 K: 1 K; 90-110 K: 2K) for 2H-Mn$_{0.28}$TaS$_2$. (b) Temperature-dependent normalized slopes $NS = S(T)/S(T_c)$ for different theoretical models. (c) Temperature-dependent spontaneous magnetization $M_s$ (left axis) and inverse initial susceptibility $\chi_0^{-1}$ (right axis) with solid fitting curves with $\mathbf{H} \parallel \mathbf{ab}$-plane. Inset shows the log$_{10}$M vs log$_{10}$H at 82.5 K with linear fitting curve. (d) Kouvel-Fisher plots of $M_s(dM_s/dT)^{-1}$ (left axis) and $\chi_0^{-1}(d\chi_0^{-1}/dT)^{-1}$ (right axis) with solid fitting curves. (e) Scaling plots of $m$ vs $h$ with the renormalized magnetization $m\equiv\varepsilon^{-\beta}M(H,\varepsilon)$ and the renormalized field $h\equiv\varepsilon^{-(\beta+\gamma)}H$ below and above $T_c$, respectively. (f) The rescaling plots of the $M(H)$ curves by $MH^{-1/\delta}$ vs $\varepsilon H^{-1/(\beta\delta)}$. (g) The calculated magnetic entropy change $-\Delta S_M(T)$ at various field changes. (h) Field dependence of the maximum magnetic entropy change $-\Delta S_M^{max}$ and the relative cooling power RCP with power law fitting in solid lines.}
\label{MTH}
\end{figure*}

Figure 1(a,b) exhibits the crystal structure and XRD 2theta scan in logarithmic scale at room temperature for  2H-Mn$_{0.28}$TaS$_2$ single crystal. The observed sharp peaks can be indexed with (00l) planes, indicating the high quality of crystal and that the plate surface of crystal is normal to the $\mathbf{c}$-axis. The lattice parameter $c$ = 12.66(2) {\AA} can be extracted by using the Bragg's law, close to the value of 12.62 {\AA} in ref.\cite{Hinode}. Figure 1(c) exhibits the normalized Mn $K$-edge XANES spectroscopy. The threshold energy $E_0$ = 6545 eV obtained from the peak of derivative curve [inset in Fig. 2(c)] indicates a mixed $2+$ and $3+$ valence but close to Mn$^{2+}$ state \cite{Subias}. Figure 1(d) exhibits the Fourier transform magnitudes of EXAFS. In a single-scattering approximation, the EXAFS could be described by \cite{Prins}:
\begin{align*}
\chi(k) = \sum_i\frac{N_iS_0^2}{kR_i^2}f_i(k,R_i)e^{-\frac{2R_i}{\lambda}}e^{-2k^2\sigma_i^2}sin[2kR_i+\delta_i(k)],
\end{align*}
where $N_i$ is the number of neighbouring atoms at a distance $R_i$ from the photoabsorbing atom. $S_0^2$ is the passive electrons reduction factor, $f_i(k, R_i)$ is the backscattering amplitude, $\lambda$ is the photoelectron mean free path, $\delta_i$ is the phase shift, and $\sigma_i^2$ is the correlated Debye-Waller factor measuring the mean square relative displacement of the photoabsorber-backscatter pairs. The main peak corresponds well to the six nearest-neighbor Mn-S [2.50(1) {\AA}] and then two next-nearest Mn-Ta [3.12(11) {\AA}] in the Fourier transform magnitudes of EXAFS [Fig. 1(d)], extracted from the model fit with fixed coordination number and $\sigma^2 = 0.006$ {\AA}$^2$. The peaks above 3.5 {\AA} are due to the longer bond distances and the multiple scattering behavior. Local crystal structure environment of Mn atom shows that Mn is well intercalated in the vdw gap of 2H-TaS$_2$ crystal.

Figure 1(e) shows the temperature-dependent magnetic susceptibility measured in H = 1 kOe applied in $\mathbf{ab}$-plane and along $\mathbf{c}$-axis with zero-field cooling (ZFC) and field cooling (FC) processes, respectively. A sharp upturn was observed for both field directions when temperature decreases, suggesting a PM-FM transition. {\color{blue}The ZFC and FC data overlap well below $T_c$, for both orientations,} indicating the high quality of single crystal. The $\chi(T)$ is nearly isotropic at high temperature, while significant anisotropy is observed at low temperature. The value of $\chi(T)$ for $\mathbf{H} \parallel \mathbf{ab}$-plane is much larger than that for $\mathbf{H} \parallel \mathbf{c}$-axis at base temperature, indicating that the magnetic moments of Mn ions tend to be arranged in the $\mathbf{ab}$ plane. The $1/\chi(T)$ from 150 to 300 K can be well fitted by the Curie-Weiss law $\chi = \chi_0 + C/(T-\theta)$ [right axis in Fig. 2(e)], where $\chi_0$ is a temperature-independent term, $C$ and $\theta$ are the Curie-Weiss constant and Weiss temperature, respectively. The obtained Weiss temperature of $\theta_{ab}$ = 102(1) K and $\theta_c$ = 100(1) K for $\mathbf{H} \parallel \mathbf{ab}$-plane and $\mathbf{H} \parallel \mathbf{c}$-axis, respectively, the positive values confirming dominance of FM exchange interactions in 2H-Mn$_{0.28}$TaS$_2$ single crystal. The derived effective moment $P_{\textrm{eff}}$ of 5.40(2) $\mu_\textrm{B}$/Mn for $\mathbf{H} \parallel \mathbf{ab}$-plane and 5.35(2) $\mu_\textrm{B}$/Mn for $\mathbf{H} \parallel \mathbf{c}$-axis, which are intermediate in value between the spin-only moments of 4.9 $\mu_\textrm{B}$ for Mn$^{3+}$ and 5.92 $\mu_\textrm{B}$ for Mn$^{2+}$ \cite{Hinode}, in line with the XANES analysis. Then we estimate the Rhodes-Wohlfarth ratio (RWR) for 2H-Mn$_{0.28}$TaS$_2$, which is defined as $P_c/P_s$ with $P_c$ calculated from $P_c(P_c+2) = P_{eff}^2$ and $P_s$ $\sim$ 4.38(1) $\mu_\textrm{B}$ is the saturation moment \cite{Wohlfarth,Moriya}. RWR is 1 for a localized system and is larger in an itinerant system. Here we obtain the RWR $\approx$ 1.03(1), indicating a localized character \cite{Motizuki}. The mechanism of FM coupling in 2H-Mn$_{0.28}$TaS$_2$ is the Ruderman-Kittel-Kasuya-Yosida (RKKY) interaction in which the local spins of intercalated Mn ions align ferromagnetically through the itinerant Ta 5d electrons \cite{Motizuki}. Figure 1(f) presents the field-dependent magnetization at 2 K. When $\mathbf{H} \parallel \mathbf{ab}$-plane, the magnetization increases sharply at low field and saturates at a very low field $\sim$ 0.2 T, confirming the easy $\mathbf{ab}$-plane. No obvious hysteresis loop (coercive field $H_c$ $<$ 10 Oe) was observed, indicating a soft in-plane FM character.

\begin{table*}
\caption{\label{tab1}Comparison of critical exponents of 2H-Mn$_{0.28}$TaS$_2$ with different theoretical models. MAP, KFP, and CI represent the modified Arrott plot, the Kouvel-Fisher plot, and the critical isotherm, respectively.}
\begin{ruledtabular}
\begin{tabular}{llllllll}
   & Reference & Technique & $T_{c-}$ & $T_{c+}$ & $\beta$ & $\gamma$ & $\delta$ \\
  \hline
  2H-Mn$_{0.28}$TaS$_2$ & This work & MAP & 82.3(1) & 82.3(1) & 0.54(1) & 1.30(3) & 3.41(1)\\
  & This work & KFP & 82.3(1) & 82.4(2) & 0.57(1) & 1.27(3) & 3.23(1) \\
  & This work & CI  &   & &   &   & 3.51(1) \\
  2D Ising & \cite{Arrott2} & Theory & & & 0.125 & 1.75 & 15.0 \\
  Mean field & \cite{Stanley} & Theory & & & 0.5 & 1.0 & 3.0 \\
  3D Heisenberg & \cite{Stanley} & Theory & & & 0.365 & 1.386 & 4.8 \\
  3D Ising & \cite{Stanley} & Theory & & & 0.325 & 1.24 & 4.82 \\
  3D XY & \cite{Fisher} & Theory & & & 0.345 & 1.316 & 4.81 \\
  Tricritical mean field & \cite{Lin} & Theory & & & 0.25 & 1.0 & 5.0
\end{tabular}
\end{ruledtabular}
\end{table*}

Figure 1(g,h) exhibits the temperature dependence of in-plane resistivity $\rho(T)$ and thermopower $S(T)$ of 2H-Mn$_{0.28}$TaS$_2$ in 0 and 9 T. Above 150 K, the linear temperature dependence of $\rho(T)$ and $S(T)$ is due to the dominant electron-phonon scattering, similar to 2H-Mn$_{0.33}$TaS$_2$ \cite{Zhang1}. With decreasing temperature, both curves show distinct changes in behavior around $T_c$, indicating that there is a considerable interaction between transport carriers and local moments. The positive values of $S(T)$ indicates that hole-type carriers dominate. As is well known, the thermopower $S(T)$ depends sensitively on the Fermi surface. The $S(T)$ gradually deviates from the linear $T$-dependence below 150 K, and presents an obvious upturn near $T_c$ as well as a broad peak around 35(5) K, as shown in Fig. 1(h), reflecting the reconstruction of Fermi surface passing through the magnetic transition and possible phonon- or magnon-drag effect at low temperature.

To obtain a precise $T_c$ and the nature of PM-FM transition in 2H-Mn$_{0.28}$TaS$_2$ single crystal, we measured dense magnetization isotherms with $\mathbf{H} \parallel \mathbf{ab}$-plane [Fig. 2(a)]. For a second-order phase transition, the critical behavior can be depicted by a series of interrelated critical exponents. Isotherms can be analyzed with the Arrott-Noakes equation of state \cite{Arrott2},
\begin{equation}
(H/M)^{1/\gamma} = a\varepsilon+bM^{1/\beta},
\end{equation}
where $\varepsilon = (T-T_c)/T_c$ is the reduced temperature, and $a$ and $b$ are constants. The $\beta$ and $\gamma$ are critical exponents that are associated with the spontaneous magnetization $M_s$ below $T_c$ and the inverse initial susceptibility $\chi_0^{-1}$ above $T_c$, respectively \cite{Stanley,Fisher,Lin}:
\begin{equation}
M_s (T) = M_0(-\varepsilon)^\beta, \varepsilon < 0, T < T_c,
\end{equation}
\begin{equation}
\chi_0^{-1} (T) = (h_0/m_0)\varepsilon^\gamma, \varepsilon > 0, T > T_c,
\end{equation}
\begin{equation}
M = DH^{1/\delta}, T = T_c,
\end{equation}
where $\delta$ is another critical exponent associated with the M(H) at $T_c$. The $M_0$, $h_0/m_0$ and $D$ are the critical amplitudes. In the critical region, right values of $\beta$, $\gamma$, and $\delta$ should generate a set of parallel straight lines of $M^{1/\beta}$ vs $(H/M)^{1/\gamma}$ in high field region. The normalized slope $NS = S(T)/S(T_c)$ with $S(T) = dM^{1/\beta}/d(H/M)^{1/\gamma}$ enables us to identify the most suitable model by comparing $NS$ with the ideal value of 1 [Fig. 2(b)]. As we can see, the $NS$ of mean-field model is mostly close to 1 below $T_c$, and it overlaps with 3D Heisenberg and 3D XY models within experimental errors above $T_c$, but much better than 2D Ising and tricritical mean-field models, indicating a clear 3D critical behavior in 2H-Mn$_{0.28}$TaS$_2$. Figure 2(c) presents the extracted $M_s(T)$ and $\chi_0^{-1}(T)$ as a function of temperature by using a rigorous iterative method \cite{Kellner,Pramanik}. According to Eqs. (2) and (3), the $\beta = 0.54(1)$, $\gamma = 1.30(3)$, and $T_c = 82.3(1)$ K, are derived. The $\delta$ can be determined directly from the inverse slope of the critical isotherm in log-scale taking into account that $M = DH^{1/\delta}$. A linear fit of log$_{10}M$ vs log$_{10}H$ at $T_c = 82.5$ K results in $\delta$ = 3.51(1) [inset in Fig. 2(c)], which is very close to the value of 3.41(1) calculated from the Widom scaling relation $\delta = 1+\gamma/\beta$ \cite{Widom}. In the Kouvel-Fisher relation \cite{Kouvel}:
\begin{equation}
M_s(T)[dM_s(T)/dT]^{-1} = (T-T_c)/\beta,
\end{equation}
\begin{equation}
\chi_0^{-1}(T)[d\chi_0^{-1}(T)/dT]^{-1} = (T-T_c)/\gamma.
\end{equation}
Linear fittings to the plots of $M_s(T)[dM_s(T)/dT]^{-1}$ and $\chi_0^{-1}(T)[d\chi_0^{-1}(T)/dT]^{-1}$ in Fig. 2(d) yield $\beta = 0.57(1)$ with $T_c = 82.4(2)$ K, and $\gamma = 1.27(3)$ with $T_c = 82.3(1)$ K, in agreement with the values obtained from the modified Arrott plot [Table I]. The $\beta$ for 2D magnets should be located in a universal window $\sim$ $0.1 \leq \beta \leq 0.25$ \cite{Taroni}. That is to say, the critical exponents of 2H-Mn$_{0.28}$TaS$_2$ exhibit apparent 3D critical phenomenon.

\begin{figure*}
\centerline{\includegraphics[scale=1]{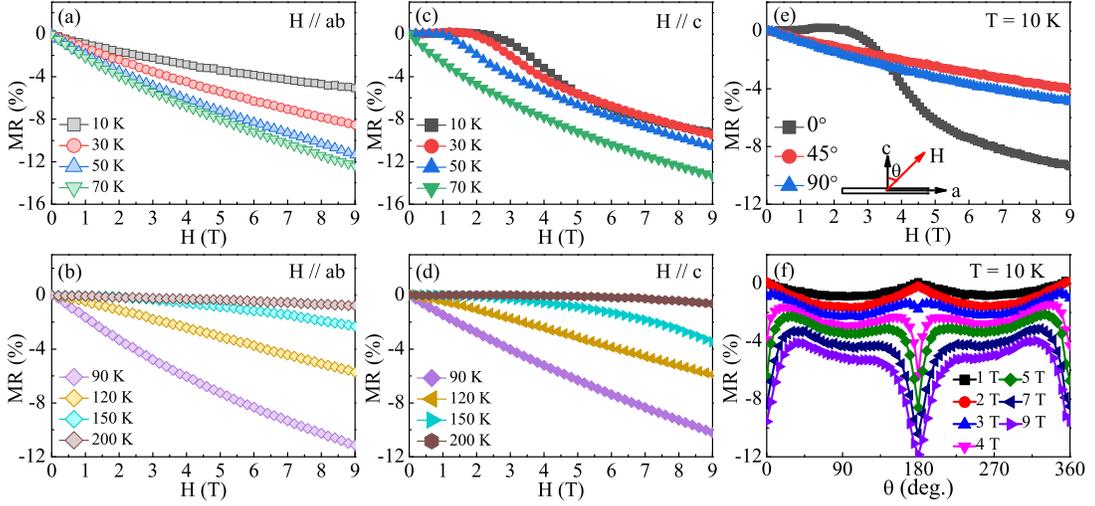}}
\caption{(Color online) Field-dependent magnetoresistance (MR) with current injected in the $\mathbf{ab}$-plane for (a,b) $\mathbf{H} \parallel \mathbf{ab}$-plane and (c,d) $\mathbf{H} \parallel \mathbf{c}$-axis at various temperatures, respectively. (e,f) Angle-dependent MR measured at $T$ = 10 K for 2H-Mn$_{0.28}$TaS$_2$.}
\label{renomalized}
\end{figure*}

Scaling analysis can be used to estimate the reliability of the obtained $\beta$, $\gamma$, $\delta$, and $T_c$. The magnetic equation of state in the critical region is expressed as \cite{Stanley}:
\begin{equation}
M(H,\varepsilon) = \varepsilon^\beta f_\pm(H/\varepsilon^{\beta+\gamma}),
\end{equation}
where $f_+$ for $T>T_c$ and $f_-$ for $T<T_c$, respectively, are the regular functions. Eq. (7) can be further written in terms of scaled magnetization $m\equiv\varepsilon^{-\beta}M(H,\varepsilon)$ and scaled field $h\equiv\varepsilon^{-(\beta+\gamma)}H$ as $m = f_\pm(h)$. This suggests that for true scaling relations and the right choice of $\beta$, $\gamma$, and $\delta$, scaled $m$ and $h$ will fall on universal curves above $T_c$ and below $T_c$, respectively. The scaled $m$ vs $h$ curves are plotted in Fig. 2(e). Obviously the lines are seperated into two branches below and above $T_c$, respectively. This can be further verified by another form of scaling equation of state \cite{Stanley},
\begin{equation}
\frac{H}{M^\delta} = k\left(\frac{\varepsilon}{H^{1/\beta}}\right),
\end{equation}
where $k(x)$ is the scaling function. According to Eq. (8), the plot of $MH^{-1/\delta}$ vs $\varepsilon H^{-1/(\beta\delta)}$ should correspond to one universal curve \cite{Phan}, which is indeed seen in Fig. 2(f), the $T_c$ locates at the zero point of the horizontal axis. The well-rescaled curves ensure that the obtained critical exponents and $T_c$ are reliable and intrinsic. Figure 2(g) presents the derived magnetic entropy change $-\Delta S_M = \int_0^H \left[\partial M(T,H)/\partial T\right]_HdH$ (left axis) \cite{Amaral}. It shows a broad peak centered near $T_c$ and the peak value monotonically increases with increasing field, reaches 1.70 J kg$^{-1}$ K$^{-1}$ in 5 T. The relative cooling power $RCP = -\Delta S_M^{max} \times \delta T_{\textrm{FWHM}}$ is also plotted (right axis), where $-\Delta S_M^{max}$ is the maximum entropy change near $T_c$ and $\delta T_{\textrm{FWHM}}$ is the full-width at half maxima \cite{Gschneidner}. The field dependence of $-\Delta S_M^{max}$ and $RCP$ at $T_c$ shows a power law $-\Delta S_M^{max} = aH^b$ and $RCP = cH^d$ \cite{VFranco}, where $c$ and $d$ are related to the critical exponents as $b = 1+(\beta-1)/(\beta+\gamma)$ and $d = 1+1/\delta$, respectively. Fitting of $-\Delta S_M^{max}$ and $RCP$ gives $b = 0.79(1)$ and $d = 1.00(2)$, which are close to the values of $b = 0.75(2)$ and $d = 1.29(1)$ calculated from the obtained $\beta$, $\gamma$, and $\delta$. This can avoid the multi-step nonlinear fitting induced deviation in the modified Arrott plot and Kouvel-Fisher plot, further verify the reliability and intrinsic of obtained critical exponents.

Then we discuss the nature as well as the range of magnetic interaction in 2H-Mn$_{0.28}$TaS$_2$. In renormalization group theory analysis the interaction decays with distance $r$ as $J(r) \approx r^{-(3+\sigma)}$, where $\sigma$ is a positive constant \cite{Fisher1972}. The susceptibility exponent $\gamma$ is:
\begin{multline}
\gamma = 1+\frac{4}{d}\left(\frac{n+2}{n+8}\right)\Delta\sigma+\frac{8(n+2)(n-4)}{d^2(n+8)^2}\\\times\left[1+\frac{2G(\frac{d}{2})(7n+20)}{(n-4)(n+8)}\right]\Delta\sigma^2,
\end{multline}
where $\Delta\sigma = (\sigma-\frac{d}{2})$ and $G(\frac{d}{2})=3-\frac{1}{4}(\frac{d}{2})^2$, $n$ is the spin dimensionality \cite{Fischer}. When $\sigma > 2$, the Heisenberg model is valid and $J(r)$ decreases faster than $r^{-5}$. When $\sigma \leq 3/2$, the mean-field model is satisfied and $J(r)$ decreases slower than $r^{-4.5}$. Here it is found that \{$d:n$\} = \{3 : 3\} and $\sigma = 1.85$ give the exponents mostly close to the experimental values in 2H-Mn$_{0.28}$TaS$_2$. Furthermore, we obtain the correlation length critical exponent $\nu$ = 0.697 ($\nu = \gamma/\sigma$, $\xi = \xi_0 |(T-T_c)/T_c|^{-\nu}$), and $\alpha$ = -0.092 ($\alpha= 2 - \nu d$), also close to the theoretical value (-0.12) for 3D Heisenberg model \cite{Le,Fisher1974}. Having delineated salient features of magnetic state, we now turn to investigation of electronic transport in magnetic field.

Figure 3(a-d) exhibits the MR measured with $\mathbf{H} \parallel \mathbf{ab}$-plane and $\mathbf{H} \parallel \mathbf{c}$-axis, respectively, at various temperatures with the current injected in the $\mathbf{ab}$-plane. When $\mathbf{H} \parallel \mathbf{ab}$-plane, the values of MR are negative in the whole temperature range and reach a maximum $\sim$ 12.4\% at 70 K, which is normally expected in an in-plane FM thin crystal \cite{L1,L2}. It is very interesting that the MR features a plateau or shoulder in low field with positive values at low temperature [Fig. 3(c)]. The maximum $\sim$ 13.4\% is observed with $\mathbf{H} \parallel \mathbf{c}$-axis at 70 K, and the field-dependent MR exhibits a similar tendency compared with those in $\mathbf{H} \parallel \mathbf{ab}$-plane when $T\geq$ 70 K. This unusual plateau feature is more apparent at lower temperature, which might be associated with a novel field-induced magnetic structure. A similar behavior was previously observed in helical magnetic MnSi nanowire due to field-induced conical or skyrmion state \cite{Du}. This unusual anisotropic MR is also plotted in Fig. 3(e) with different field orientations. A detailed angle-dependent MR under different fields at 10 K was further measured [Fig. 3(f)]. The angle-dependent MR shows an obvious two-fold symmetry at 1 T and 2 T, however an extra peak shows up when $\mathbf{H}\parallel \mathbf{c}$-axis at 3 T, which grows rapidly and takes over the maximum above 4 T. Hence, a field-induced magnetic structure is likely in 2H-Mn$_{0.28}$TaS$_2$ single crystals which calls further in-depth investigation as well as its thickness-dependent properties.

\section{CONCLUSIONS}

In summary, we systematically studied the mechanism of the PM-FM transition and magnetotransport in 2H-Mn$_{0.28}$TaS$_2$ single crystals. The spin interaction in 2H-Mn$_{0.28}$TaS$_2$ is close to a 3D Heisenberg model coupled with attractive long-range exchange interaction decaying with distance as $J(r)\approx r^{-4.85}$. A field-induced novel magnetic structure is proposed based on the unusual angle-dependent MR behavior. It is also of high interest to explore the magnetic properties of 2H-Mn$_{0.28}$TaS$_2$ at the 2D limit down to a monolayer in future studies due to its large magnetic anisotropy. {\color{blue}The exfoliation of thin samples of such intercalated transition metal dichalcogenides can refer to the method in Ref. \cite{Danz}.}

\section*{Acknowledgements}

Work at BNL is supported by the Office of Basic Energy Sciences, Materials Sciences and Engineering Division, U.S. Department of Energy (DOE) under Contract No. DE-SC0012704. This research used the 8-ID beamline of the NSLS II, a U.S. DOE Office of Science User Facility operated for the DOE Office of Science by BNL under Contract No. DE-SC0012704.\\

$^{*}$Present address: Los Alamos National Laboratory, MS K764, Los Alamos NM 87545, USA\\

$^{\dag}$Present address: ALBA Synchrotron Light Source, Cerdanyola del Valles, E-08290 Barcelona, Spain.

\end{document}